%% file: cr_main.tex
\documentclass[10pt,twocolumn,letterpaper]{article}

\usepackage{colortbl}
\usepackage{iccv}
\usepackage{times}
\usepackage{epsfig}
\usepackage{graphicx}
\usepackage{amsmath}
\usepackage{amssymb}
\input{math_commands.tex}

\usepackage[noadjust]{cite}
\usepackage{url}
\usepackage{wrapfig}
\usepackage{lipsum}
\usepackage{diagbox}

\usepackage{tikz}
\usepackage{amsmath}
\usepackage{epsfig,amsmath,amsfonts,multirow,makecell,caption,soul,csquotes,color,wrapfig,subcaption,mathtools,bm,spverbatim,booktabs,xcolor,amsthm,tcolorbox,wrapfig,enumitem}

\usepackage{filecontents}
\usepackage[accsupp]{axessibility}  

\usepackage[pagebackref=true,breaklinks=true,letterpaper=true,colorlinks,bookmarks=false]{hyperref}

\newcommand{\system}{SysAdv\xspace}

\newcommand{\RNum}[1]{\uppercase\expandafter{\romannumeral #1\relax}}

\newcommand{\meq}[1]{Eq.\,(\ref{#1})}

\usepackage[breaklinks=true,bookmarks=false]{hyperref}
\usepackage{marvosym}

\iccvfinalcopy 

\def\iccvPaperID{****} 

\ificcvfinal\fi

\begin{document}

\title{Does Physical Adversarial Example Really Matter to Autonomous Driving? Towards System-Level Effect of Adversarial Object Evasion Attack}

\author{Ningfei Wang \quad Yunpeng Luo \quad Takami Sato \quad Kaidi Xu$^\dagger$ \quad Qi Alfred Chen \\
    University of California, Irvine \ \ \  \{ningfei.wang, yunpel3, takamis, alfchen\}@uci.edu\\
    $^\dagger$Drexel University  \ \ \ kx46@drexel.edu
    }

\maketitle
\ificcvfinal\thispagestyle{empty}\fi

\input{src/abstract}

\input{src/intro}

\input{src/background}

\input{src/measurement}
\input{src/observation_improvement}

\input{src/evaluation}

\input{src/discussion}
\input{src/limitation}

\input{src/conclusion}

\input{src/ack}

{\small
\bibliographystyle{ieee_fullname}
\bibliography{main}
}

\end{document}

%% file: math_commands.tex
\usepackage{amsmath,amsfonts,bm}

\def\eqref#1{equation~\ref{#1}}

\def\1{\bm{1}}

\DeclareMathAlphabet{\mathsfit}{\encodingdefault}{\sfdefault}{m}{sl}
\SetMathAlphabet{\mathsfit}{bold}{\encodingdefault}{\sfdefault}{bx}{n}

\def\gL{{\mathcal{L}}}

\DeclareMathOperator*{\argmin}{arg\,min}

%% file: src/abstract.tex
\begin{abstract}
In autonomous driving (AD), accurate perception is indispensable to achieving safe and secure driving. Due to its safety-criticality, the security of AD perception has been widely studied. Among different attacks on AD perception, the physical adversarial object evasion attacks are especially severe. However, we find that all existing literature only evaluates their attack effect at the targeted AI component level but not \textit{at the system level}, i.e., with the entire system semantics and context such as the full AD pipeline. Thereby, this raises a critical research question: can these existing researches effectively achieve system-level attack effects (e.g., traffic rule violations) in the real-world AD context? In this work, we conduct the first measurement study on whether and how effectively the existing designs can lead to system-level effects, especially for the STOP sign-evasion attacks due to their popularity and severity. Our evaluation results show that all the representative prior works cannot achieve any system-level effects. We observe two design limitations in the prior works: 1) physical model-inconsistent object size distribution in pixel sampling and 2) lack of vehicle plant model and AD system model consideration. Then, we propose \system, a novel system-driven attack design  in the AD context and our evaluation results show that the system-level effects can be significantly improved, i.e., the violation rate increases by around 70\%.
\end{abstract}

%% file: src/intro.tex
\vspace{-0.5cm}
\section{Introduction}
\label{sec:intro}
Autonomous Driving (AD) vehicles are now a reality in our daily life, where a wide variety of commercial and private AD vehicles are driving on the road. For instance, the millions of Tesla cars~\cite{Kane2021Tesla} equipped with Autopilot~\cite{Tesla2022autopilot} are publicly available. To ensure safe and correct driving, a fundamental pillar is \textit{perception}, which is designed to detect surrounding objects in real time. Due to the safety- and security-criticality of AD perception, various prior works have studied its security, especially the ones that aim at causing the evasion of critical physical road objects (e.g., STOP signs and pedestrians), or \textit{physical adversarial object evasion attack}~\cite{jia2022fooling, xu2020adversarial, chen2018shapeshifter, wu2020making, zhao2019seeing, sp:2021:ningfei:msf-adv, eykholt2018physical, lovisotto2021slap, cao20203d}.

Although these attacks are all motivated by causing erroneous driving behaviors at the AD system level (e.g., vehicle collisions and traffic rule violations), we find that so far they predominately only evaluate the attack success \textit{at the targeted AI component level alone} (e.g., judged by per-frame object misdetection rates~\cite{chen2018shapeshifter, eykholt2018physical, xu2020adversarial, zhao2019seeing, jia2022fooling}), without further evaluation \textit{at the system level}. Specifically, to systematically perform such system-level evaluation, we need to measure the end-to-end system-level attack success metrics (e.g., collision rates) with the full system-level attack context enclosing the attack-targeted AI component, for example, the remaining AD system pipeline such as object tracking, planning, and control, closed-loop control, and the attack-targeted driving scenario. In this paper, we call such system-level attack context \textit{system model} for such adversarial attacks (\S\ref{sec:background-system-model}).
This thus raises a critical research question: \textit{can these existing works on physical adversarial object evasion attacks effectively achieve the desired system-level attack effects in the realistic AD system settings?}

To systematically answer this critical research question, we conduct the first measurement study on representative prior object-evasion attacks with regard to their capabilities in causing system-level effects (\S\ref{sec:measurement}). We propose a general framework, i.e., a system model, including perception modeling from the physical world, to measure
STOP sign-evasion attack which is our target due to its high representativeness~\cite{shen2022sok} and its direct impacts on driving correctness and road safety. 
Our results show that \textit{all} the representative existing works cannot cause any STOP sign traffic rule violation within the system model including a representative closed-loop control AD system in the common speed range for STOP sign-controlled roads in the real world even though the most effective attack can achieve more than 
70\% average attack success rate at the AI component alone.

We further investigate the root causes and find that all the existing works have design limitations on achieving effective system-level effects due to the lack of a system model in AD context for attack design: 1) physical model-inconsistent object size distribution in pixel sampling and 2) lack of vehicle plant model and AD system model consideration (detailed in~\S\ref{sec:design-lh-improve-proposal}). We further propose \system, a system-driven attack design, which can be integrated with all state-of-the-art attack methods to significantly improve system-level effects by overcoming the two limitations.

We evaluate our novel proposed attack design in our platform and show that the system-level effect can be significantly improved in~\S\ref{sec:hypothesis-val}, i.e., the system violation rate can be increased by around 70\%. To further validate the generality of our
attack, we also examine generality on different AD system parameters (\S\ref{sec:general_ad_settings}) and different object types (\S\ref{sec:general_object_type}), which shows improvement at both component- and system-level.
Demo videos are at the project website: \textbf{\url{https://sites.google.com/view/cav-sec/sysadv}}.

To sum up, this paper makes the following contributions:

\begin{itemize}
\vspace{-0.3cm}
  \item We conduct the first measurement study on the system-level effect of the representative prior object-evasion attacks with our proposed novel evaluation framework (i.e., system model) including 4 popular object detectors and 3 state-of-the-art object-evasion attacks.
  \vspace{-0.1cm}
  \item We identify the limitations of prior works which hinder them in potently achieving system-level effects and propose \system, a system-driven adversarial object-evasion attack with the system model in AD context.
  \item We further evaluate \system
  and show that the system-level effect of \system can be significantly improved, i.e., the system violation rate increases by around 70\%.

\end{itemize}

%% file: src/background.tex
\section{Related Work and Background}
\label{sec:background}
\textbf{Camera-based AD perception.}
\label{sec:background-camera}
Camera-based AD perception generally leverages DNN-based object detection to detect or recognize road objects of various categories (e.g., traffic signs, vehicles, and pedestrians) in consecutive image frames~\cite{carranza2020performance}. State-of-the-art DNN-based object detectors can be classified into two categories: one-stage object detector, and two-stage object detector~\cite{zou2023object}. The former, such as YOLO~\cite{redmon2017yolo9000, redmon2018yolov3, Jocher2022yolov5}, usually has higher detection speed, while the latter, such as Faster R-CNN~\cite{ren2015faster}, usually has higher detection accuracy. 
In this paper, we focus on the security aspects of camera-based AD perception and perform the corresponding experiments on both object detector categories.
We perform the measurement study of physical adversarial object evasion attack in AD perception~\S\ref{sec:measurement} including these two kinds of object detectors.

\textbf{Physical adversarial object evasion attacks in AD context.}
\label{sec:background-adversarial}
Recent works find that DNN models are generally vulnerable to adversarial attacks~\cite{goodfellow:fsgm, carlini:cw, mkadry2017towards, xiao2019meshadv, zhang2020interpretable, ma2023wip}. Due to the direct reliance of camera-based AD perception on DNN object detectors, various prior works have explored the feasibility of adversarial attacks in AD context~\cite{jia2022fooling, zhao2019seeing, xu2020adversarial, zolfi2021translucent, wang2021can, shen2022sok, wang2022poster, dipalma2021security, ma2023wip, sato2020hold, luo2022infrastructure, sato2021wip}. Among them, \textit{physical adversarial object evasion attacks}, which typically use physical-world attack vectors such as malicious patches to cause the disappearance of road objects (e.g., pedestrians and traffic signs)~\cite{jia2022fooling, eykholt2018physical,zhao2019seeing, xu2020adversarial, chen2018shapeshifter, wu2020making}, are especially severe due to their direct impacts on driving correctness and road safety. However, as detailed in later sections, we find that so far the considerations and integration of the corresponding \textit{system models} (detailed below) in the prior works are \textit{far from enough} in both attack designs and evaluation, which substantially jeopardizes the meaningfulness of their designs from the end-to-end AD driving perspective (\S\ref{sec:measurement}).

{\bf Gap between AI component errors and their system-level effect.} We do not intend to claim to be the first to point out, analyze, measure, or optimize the gap between AI component errors and their system-level effect in general; there exists a large body of prior works in various other problem contexts (e.g., computer vision system~\cite{jain1991ignorance, ramesh1997computer, ji1999error},  image analysis~\cite{haralick1992performance, zhang2018visual}, camera surveillance~\cite{greiffenhagen2000statistical, greiffenhagen2001systematic}, video analytics~\cite{thacker2008performance, greiffenhagen2001design}, planning~\cite{phillips2021deep, philion2020learning, topan2022interaction}, and control~\cite{topan2022interaction}) across academia and industry that have studied the characterization and/or optimization of end-to-end system performance~\cite{gog2021pylot, caesar2020nuscenes} with regard to AI/vision component errors. Nevertheless, to the best of our knowledge, none of them 1) quantified such gaps in the context of adversarial attacks on autonomous systems, especially those in real-world system setups; and 2) identified novel designs that can systematically address or fill such gaps on autonomous systems, which we believe are our novel and unique contributions.

\textbf{Systems model for AD AI adversarial attacks.}
\label{sec:background-system-model}
To understand the end-to-end system-level impacts of an adversarial attack against a targeted AI component in an AD system (e.g., whether it can indeed effectively cause undesired AD system-level property violations), we need to systematically consider and integrate the overall system semantics and context that enclose such AI component into the security analysis~\cite{verifai-cav19, seshia2022toward}. In this paper, we call a systematic abstraction of such system semantics and context the \textit{system model} of such AD AI adversarial attacks. Specifically, in the AD context we identify 3 essential sub-components in such system model: 1) \textit{the AD system model}, i.e., the full-stack AD system pipeline that encloses the attack-targeted AI components and closed-loop control, e.g., the object tracking, planning, and control pipeline for the object detection AI component; 
2) \textit{the vehicle plant model}~\cite{nese2015systematic, verifai-cav19}, which defines the physical properties of the underlying vehicle system under control, e.g., maximum/minimum acceleration/deceleration, steering rates, sensor mounting positions, etc.; and 3) \textit{the attack-targeted operation scenario model}, which defines the physical driving environment setup, driving norms (e.g., traffic rules), and the system-level attack goal (e.g., vehicle collision, traffic rule violation, etc.) targeted by the AD AI adversarial attack.

\begin{figure}[t]
    \footnotesize
      \centering 
          \includegraphics[width=\linewidth]{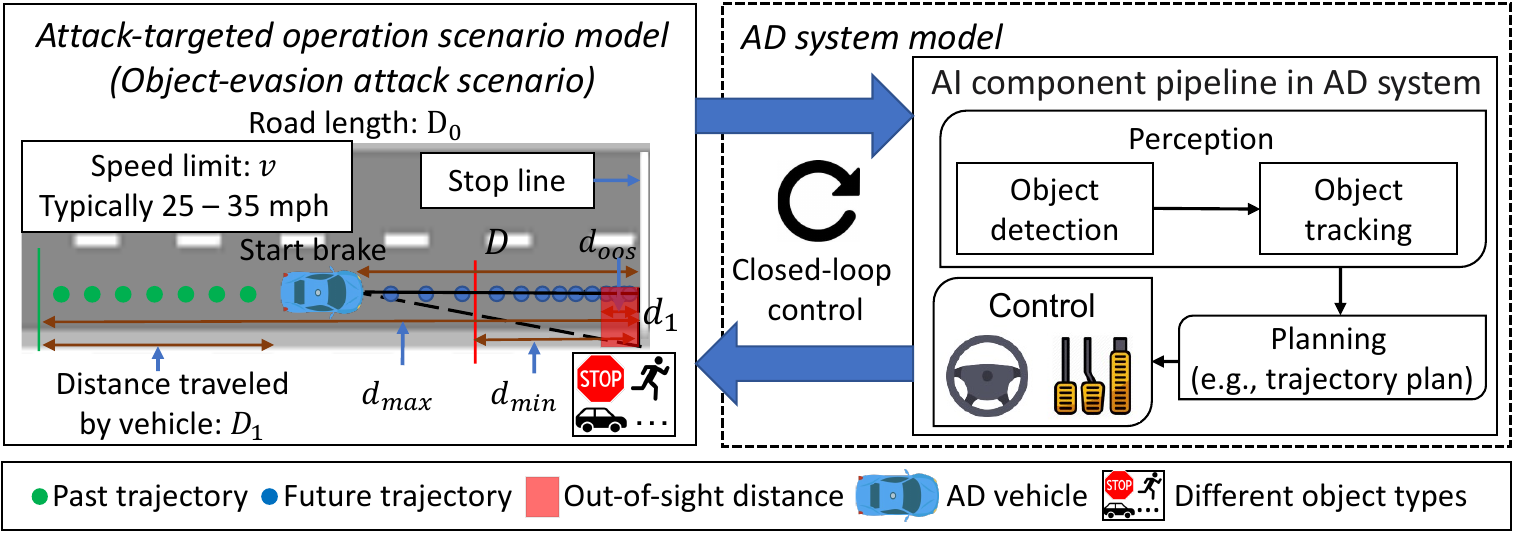}  
          \vspace{-0.5cm}
    \caption{Illustration of the system model for adversarial object-evasion attacks in AD context.}
    
        \label{fig:system_model}
        \vspace{-0.5cm}

\end{figure}

\textbf{System model for adversarial object-evasion attacks.} Fig.~\ref{fig:system_model} illustrates the aforementioned system model defined for the adversarial object-evasion attack.
The AD system model for object detection, the targeted AI component in adversarial object-evasion attacks, mainly includes its downstream tasks of object tracking, planning, and control, and closed-loop control. The vehicle plant model mainly includes the physical properties related to longitudinal control, e.g., the minimum brake distance ($d_{min}$), and the distance to the stop line (stop to avoid violating traffic rules or crashes) where the stop line is out of sight in the camera image $d_{oos}$ (depending on the hood length and the camera mounting position). The operation scenario model includes the speed limit, lane width, the relative positioning and facing of the object to the ego lane, the driving norm that the vehicle typically drives at constant speed before it starts to see the object ($d_{max}$), and the system-level attack goal that triggers the traffic rule violation (i.e., hit into the object or exceeding the stop line). We will use this system model in our studies in the following sections. There exists several example attacks for the system model such as STOP sign-evasion attack, which is the most extensively-studied physical adversarial object evasion attack in AD context~\cite{shen2022sok}, and thus will be the main focus of our study in later sections; pedestrian-evasion attack~\cite{xu2020adversarial}; car-evasion attack~\cite{wang2022fca}; etc.

%% file: src/measurement.tex
\section{System-Level Effect of Prior Works}
\label{sec:measurement}
\textbf{Scientific gap in existing works: Lack of system-level evaluation.} Despite a plethora of published attack works on physical adversarial object evasion attacks in AD context (\S\ref{sec:background-adversarial}), we find that actually \textit{all of them} only evaluate their attack effect \textit{at the targeted AI component level} (i.e., judged by per-frame object misdetection rates~\cite{eykholt2018physical, xu2020adversarial, zhao2019seeing, jia2022fooling}), without any evaluation \textit{at the system level}, i.e., with the corresponding system models for such attacks as described in~\S\ref{sec:background-system-model}. 
However, in the Cyber-Physical System (CPS) area, it is widely recognized that in AD system, AI component-level errors do not necessarily lead to system-level effects (e.g., vehicle collisions)~\cite{verifai-cav19, seshia2022toward, jia2020fooling}. 
Thus, without system-level evaluation, it can be highly difficult to scientifically know whether the attack is actually meaningful from the end-to-end AD driving perspective. We view this as a critical scientific gap in this current research space, and to address this, we perform a measurement study on the existing works about their system-level effects. 
We choose to focus on \textit{adversarial STOP sign-evasion attck} as our target due to its high representativeness in this research space and also its direct impacts on driving correctness and road safety (\S\ref{sec:background-adversarial}).

\begin{figure}[t]
    \footnotesize
      \centering
        \includegraphics[width=\linewidth]{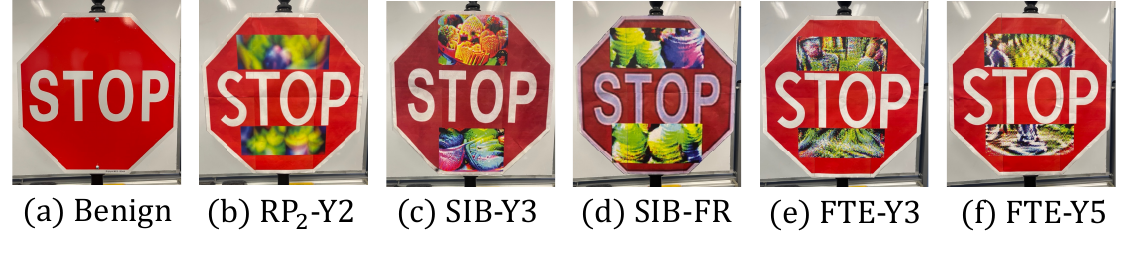}  
    \vspace{-0.7cm}
    \caption{Visualisation of STOP signs attack reproduction (in Table~\ref{tab:measurement-model-attack}) for measurement study in physical world.}
        \label{fig:stop_sign_measurement_visual}
        \vspace{-0.5cm}
\end{figure}

\subsection{Attack Formulation and Selection}
\label{sec:attack-selection}

{\bf Attack formulation.} We assume that the attacker can arbitrarily manipulate pixels within restricted regions known as adversarial patch attack~\cite{brown2017adversarial, zhao2019seeing, eykholt2018physical}. Such a patch attack is easy to deploy in the real-world and very stealthy. We consider the patch stays on the STOP sign shown in Fig.~\ref{fig:stop_sign_measurement_visual}.

{\bf Selection of prior STOP sign attack works and their reproduction.} There are various prior works on physical adversarial STOP sign-evasion attacks~\cite{lu2017no, jia2022fooling, eykholt2018physical, zhao2019seeing, xue2021naturalae,lu2017adversarial, chen2018shapeshifter}. 
To perform our system-level effect measurement, we select the most effective ones at AI component level as representative examples. Four model designs (including one-stage and two-stage object detectors in~\S\ref{sec:background}) have been covered. For each model, we select the most effective attack design published so far which are shown in Table~\ref{tab:measurement-model-attack}.
However, all the STOP sign attacks in Table~\ref{tab:measurement-model-attack} do not provide the source code. Since we tried to contact the authors of the attacks for the source code but they all cannot provide it, we try our best to reproduce some of the works. Currently, we only have the reproduction for RP$_2$ and FTE. For SIB, we directly use the STOP sign images shared by the authors of that paper used for their physical-world experiments.
We print the high-resolution STOP signs on multiple ledger-size papers and concatenate them together to form full-size real STOP signs which are shown in Fig.~\ref{fig:stop_sign_measurement_visual}.

To demonstrate the reproduction correctness, we utilize their original evaluation setups for our trials. Our results are generally similar to theirs confirming the correctness of reproduction. For instance, the original RP$_2$ paper~\cite{eykholt2018physical} reports an attack success rate of approximately 63.5\% from 0 to 30 feet.
With the same setup (outdoor), our results provide a 61.0\% attack success rate — nearly mirroring the original.
Note that SIB attack on the FR in Table~\ref{tab:measurement-results} seems anomalous: it records around 47\% attack success rate only from 40 to 45 meters, while consistently registering 0\% in others. Despite the patch being provided by the authors, the pre-trained FR can be different, where we use MMDetection~\cite{mmdetection}, a PyTorch-based object detection toolbox. Given such potential low transferability, the attack may be less effective compared to their original results. However, this is our best effort to reproduce their results faithfully.

\subsection{Measurement Methodology and Setup}
\label{sec:measurement-methodology-and-setup}
To measure system-level effects, we adopt a simulation-centric evaluation methodology, which has been widely adopted both in academia~\cite{ndss:2022:ziwen:planfuzz,sec:2021:sato:drpattack} and in industry~\cite{Waymo-Safety-Report, Scaling-Simulation} due to the inherent limitations of real-road AD testing in cost, safety, efficiency, and corner-case coverage. 
In this study, we use SVL, a production-grade high-fidelity AD simulator designed for AD systems~\cite{rong2020lgsvl}. As repeatedly demonstrated in various prior works, the end-to-end AD system-level evaluation results in SVL can highly correlate with the same setup tested in the physical world~\cite{ndss:2022:ziwen:planfuzz, sec:2021:sato:drpattack}. 
To ensure the fidelity of our evaluation results, 
we improve the fidelity of the rendering process by modeling the perception results in the real world with a practical setup (details below). 
Note that the attacks themselves are agnostic to map and time by design, and thus are not generally affected by their changes. In SVL, we use San Francisco map on a sunny day at noon, which is the most representative setup.

\begin{figure}[t]
    \footnotesize
      \centering
 \includegraphics[width=\linewidth]{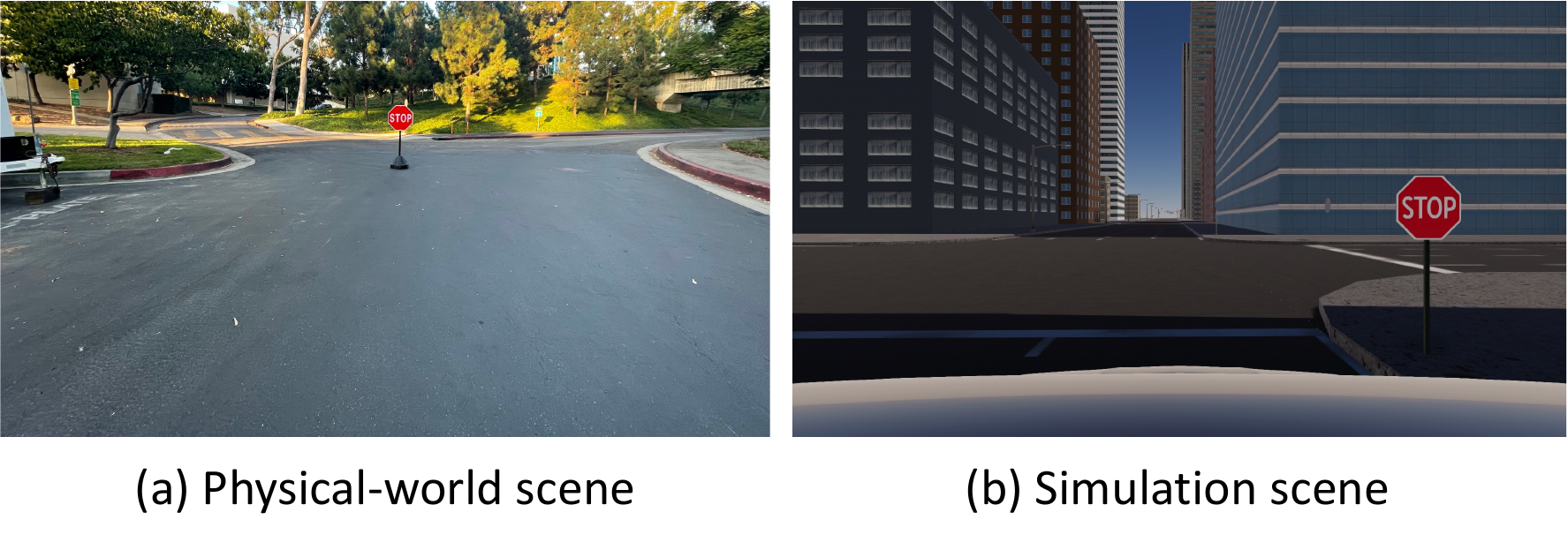}  
 \vspace{-0.6cm}
    \caption{Experiment scenes. (a) Real-world scene with real road and injected STOP sign; (b) SVL simulation scene with the San Francisco map in a sunny day at noon. }
        \label{fig:real-world-scene}
        \vspace{-0.3cm}
\end{figure}

{\bf Perception results modeling from physical world.} 
To enhance the perception fidelity of simulators, we model the perception results using a practical setup in the real world. This approach represents our best effort to improve the fidelity of the simulation due to the experimental feasibility.
Previous studies collect video frames by directly moving towards the STOP sign and simulate varying view angles by rotating the STOP sign itself. This approach is not practical since the vehicles do not directly drive towards the STOP sign, and the STOP sign should instead be located on the roadside as shown in Fig.~\ref{fig:system_model}. To improve such unrealistic setups, we follow the system model defined in~\S\ref{sec:background-camera}.
We recorded several pieces of video along the driving direction $D$ using an iPhone 12 Pro Max starting from 45 m to 4 m (4 m is the $d_{oos}$ in~\S\ref{sec:background-camera}). We choose 45 m since 1) it is the minimal brake distance for speed above 50 mph, which exceeds the usual maximum speed of STOP sign areas, and 2) it is already much larger than the maximum distance evaluated in all the prior STOP sign-evasion attack works. We separate the whole range into 9 pieces, each spanning 5 m except the one near the STOP sign, which is 1 m long. Then, we record a video in each region and feed the video into the object detectors to model the perception results. 
We perform these experiments on sunny days as shown in Fig.~\ref{fig:real-world-scene}. 
With that,
we perform perception results injection at the output of the object detection task in the AD system, i.e., first read the ground-truth STOP sign detection results from the simulator and then drop/keep the detection results based on detection rate. 
For instance, if the attack success rate is 60\% for a range, we will randomly drop the STOP sign detection results with a possibility of 60\% in that range.

\textbf{Evaluated AD system pipeline.} The AD system pipeline includes representative downstream tasks after object detection, which contains 1) a tracking step using a general Kalman Filter based multi-object tracker~\cite{luo2021multiple}; 2) a planning step using a lane-following planner from Baidu Apollo~\cite{Apollo2022baidu}, an industry-grade full-stack AD system; and 3) a control step using classic controllers, i.e., PID for longitudinal control used in OpenPilot~\cite{OpenPilot2022}, a production-grade Level-2 AD system, and Stanley~\cite{hoffmann2007autonomous} for lateral control.

{\bf Speed selection.} The driving speed is from 25 to 35 mph, with a step size of 5 mph, which is the most common speed range for STOP sign-controlled roads in the real world. 25 mph~\cite{atlantaga} is the common speed limit for the STOP sign-controlled road intersections, which is more likely to avoid a crash, and 35 mph~\cite{visitcalifornia} is the most common speed limit for city streets, which STOP signs are designed for.

\begin{table}[t]
\tabcolsep 0.01in
\footnotesize
    \caption{Selection of the representative prior works. Specifically, for each of the 4 model types targeted by prior works, we select the most effective attack design published so far.}
    \centering
    \begin{tabular}{ccccc}

    \toprule
    Model & YOLO v5 (Y5) & YOLO v3 (Y3) & YOLO v2 (Y2) & Faster RCNN (FR)\\
    
    \midrule
    Attack & FTE~\cite{jia2022fooling} & SIB~\cite{zhao2019seeing} & RP$_2$ ~\cite{eykholt2018physical} & SIB~\cite{zhao2019seeing} \\

        \bottomrule
         
    \end{tabular}

    \label{tab:measurement-model-attack}
\end{table}

\begin{table}[t]
\tabcolsep 0.01in
\footnotesize
    \caption{System-level violation rate in the simulation-based testing and component-level overall ASR for model Y2, Y3, Y5, and FR in benign and attacked scenarios. 10 runs for each cell with different initial AD position. B: benign; Sys: system; Comp: component; ASR: attack success rate. 
    }
    \centering
    \begin{tabular}{ccccccccccc}

    \toprule
    & & \multicolumn{2}{c}{Y2} & \multicolumn{3}{c}{Y3} & \multicolumn{2}{c}{Y5} & \multicolumn{2}{c}{FR}\\
    
    \cmidrule(lr){3-4}
    \cmidrule(lr){5-7}
    \cmidrule(lr){8-9}
    \cmidrule(lr){10-11}
    
  \multirow{-2}{*}[3pt]{\shortstack{Eval. \\ level}} &  \multirow{-2}{*}[3pt]{ \shortstack{Speed \\ (mph)}}& B & RP$_2$ & B & SIB & FTE & B & FTE & B & SIB\\
    
    \midrule

      Sys (violation) & 25, 30, 35 &   0\% &   0\% & 0\% & 0\% & 0\% & 0\% & 0\%& 0\% & 0\% \\
    
    \midrule
     Comp (ASR) & Overall & - & 71.2\% & - & 53.1\% & 53.3\% & - & 41.0\% & -  & 5.2\%\\
    
        \bottomrule
         
    \end{tabular}

    \label{tab:violation-existing-work}
   
\vspace{-0.5cm}
\end{table}

\begin{table*}[t]
\tabcolsep 0.07in
\footnotesize
    \caption{Detection rates of different objectors in benign, RP$_2$-, SIB-, and FTE-attacked scenarios tested in the physical world
    for perception results modeling (shown in~\S\ref{sec:measurement-methodology-and-setup}). Each detection rate below is calculated with at least 400 video frames.}
    \vspace{-0.25cm}
    \centering
    \begin{tabular}{ccccccccccc}

    \toprule
    \diagbox{Object Detector}{Distance range (m)} &  & 4 - 5 & 5 - 10 & 10 - 15 & 15 - 20 & 20 - 25 & 25 - 30& 30 - 35 & 35 - 40 & 40 - 45\\
    
    \midrule
    & Benign &  100\% &  100\% & 71.3\% & 31.3\% & 0\% & 0\% & 0\% &  0\% & 0\%  \\
   \multirow{-2}{*}{YOLO v2 (Y2)} & RP$_2$~\cite{eykholt2018physical} & 58.2\% & 90.0\% & 76.2\% & 34.6\% & 0.1\% & 0\% & 0\% & 0\% & 0\%\\
   
     \midrule
    
    & Benign &    100\% &  100\% &  100\% & 100\% &  80.1\% &  11.8\% & 6.7\% & 1.0\%  & 0\%\\
    & SIB~\cite{zhao2019seeing} & 93.7\% & 100\% & 100\% & 90.4\% & 38.2\% & 0\% & 0\% & 0\% & 0\%\\
   \multirow{-3}{*}{YOLO v3 (Y3)} & FTE~\cite{jia2022fooling} & 89.9\% & 100\% & 100\% &  87.3\% & 42.9\% & 0.6\% &  0\% &  0\% & 0\%\\
   \midrule
  
    & Benign &  100\% &  100\% & 100\% & 100\% & 98.7\% & 89.4\% & 52.3\% &  25.3\% & 0\%  \\
   \multirow{-2}{*}{YOLO v5 (Y5)} & FTE~\cite{jia2022fooling} & 91.2\% & 100\% & 100\% & 99.7\% & 88.2\% & 48.4\% & 3.9\% & 0\% & 0\%\\
   
   \midrule
  
    & Benign &  100\% & 100\% & 100\% & 100\% & 100\% &100\% &100\% & 100\% & 100\%\\
   \multirow{-2}{*}{Faster-RCNN (FR)} & SIB~\cite{zhao2019seeing} &  100\% &  100\% &  100\% & 100\% &  100\% &  100\% & 100\% & 100\%  & 53.2\% \\

        \bottomrule
         
    \end{tabular}

    \label{tab:measurement-results}
   \vspace{-0.3cm}
\end{table*}

\subsection{Measurement Results}

We evaluate the targeted AD system-level attack effect, i.e., STOP sign violation rate,
by defining the STOP sign violation rate as $\dfrac{N_\textrm{violation}}{N_\textrm{total}}$, in which $N_\textrm{violation}$ means the number of runs where the AD vehicle exceeds the stop line and $N_\textrm{total}$ is the number of total runs.
Table~\ref{tab:violation-existing-work} shows the results where each speed has 10 runs with random initialization of the AD vehicle position while the perception results modeling from real world is shown in Table~\ref{tab:measurement-results}. To our surprise, \textit{none} of the existing representative attacks can trigger STOP sign violations in \textit{any} of the common speeds for STOP sign-controlled roads when the benign performs well, though most of the attacks are effective in the component (i.e., with about 45\% average attack success rate across the 5 attacks). After inspecting the details, we find that the STOP sign is always tracked at the object tracking step before reaching the minimum brake distance of the AD vehicle due to the low attack success rate in such regions. 
Taking SIB attack on Y3 as an example, the brake distance for 30 mph is around 15 m. In the benign scenario, the detection rate for 15-20 m is 100\%, while the SIB attack can still have 90.4\% detection rate as shown in Table~\ref{tab:measurement-results}, which is not enough to make the tracking vanish before the minimum braking distance.

%% file: src/observation_improvement.tex
\section{System-Driven Attack Design}
\label{sec:design-lh-improve-proposal}
After realizing that existing works cannot provide any system-level violation in AD context,
we propose \system, a system-driven attack design, which can be integrated with all the existing attacks to improve system-level effects.

\subsection{System-Driven Attack Design Framework}
\label{sec:attack_design_overview}
For the attack design in the prior works~\cite{eykholt2018physical, zhao2019seeing, jia2022fooling, chen2018shapeshifter}, we can abstract the key part for attack generation:
\begin{equation}
\label{eq:eot-distribution}
\small
\argmin_{p_a}\ \  \mathbb{E}_{s\sim \mathcal{S}}[\gL( M(p_a, O, s, B), \gamma )]
\vspace{-0.1cm}
\end{equation}
$\mathcal{S}$ is the distribution to sample different object sizes in pixels, which is a very important factor in achieving the robust attack at different distances between the AD vehicle and the object. The $\gL$ is the loss function used in the prior attacks to achieve high attack effectiveness, $p_a$ is the adversarial patch, $O$ is the object, and function $M(p_a, O, s, B)$ indicates applying $p_a$ to $O$, then resizing object size in pixel to $s$, and applying $O$ into the background $B$, $\gamma$ means other inputs for loss function in the prior works (e.g., the bounding box information and threshold) related to the object detector alone. After investigation on \meq{eq:eot-distribution}, we find out that the system model can be involved into two parts, i.e., $\mathcal{S}$ and function $M(.)$, which do not rely on object detector alone. 

After exploring all the prior works, we discover that all of them do not consider such a system model information into their attack designs, which hinder them to achieve potent system-level effects in AD context. Thus, we propose two novel system-driven designs to significantly improve the system-level effects. Specifically, we involve the system model information into $\mathcal{S}$ and function $M(.)$ from \meq{eq:eot-distribution}.

\subsection{Physical Model-Inconsistent Object Size Distribution in Pixel Sampling}
\label{sec:dlh1}

In the prior works~\cite{zhao2019seeing, jia2022fooling, eykholt2018physical}, to make the attack robust to different distance,
Expectation over Transformation (EoT)~\cite{athalye2018synthesizing} is used to \textit{uniformly} sample the object size ($\mathcal{S}$ in \meq{eq:eot-distribution}) in a certain range~\cite{chen2018shapeshifter, jia2022fooling, athalye2018synthesizing}. However, with system model (\S\ref{sec:background-camera}), we find that this assumption is not held,
which leads to the first observation: \textit{physical model-inconsistent object size distribution in pixel sampling.} To justify the observation, we perform the experimental and theoretical analysis with STOP sign system model as an example. 

\begin{figure*}[t]
\begin{minipage}[b]{0.66\linewidth}
    \footnotesize
      \centering

        \includegraphics[width=\linewidth]{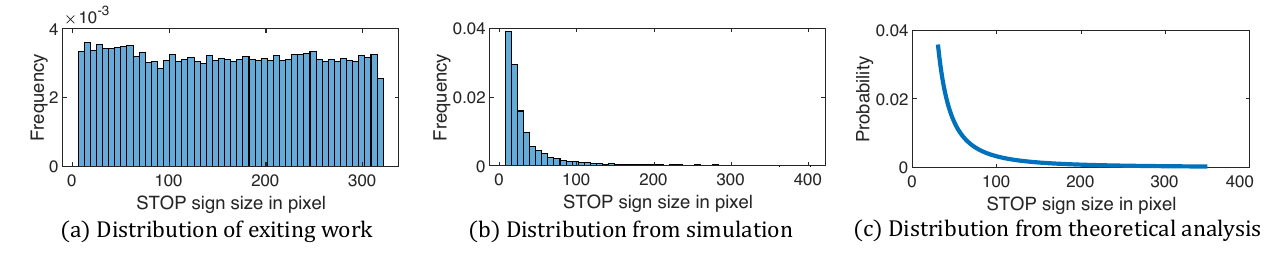}
    \vspace{-0.7cm}
    \caption{Different STOP sign size distribution: (a) state-of-the-art existing work~\cite{jia2022fooling}, (b) our experimental analysis, and (c) our theoretical analysis. }
        \label{fig:distribution}
\end{minipage}
\hfill
\begin{minipage}[b]{0.31\linewidth}
        \footnotesize
      \centering
          \includegraphics[width=\linewidth]{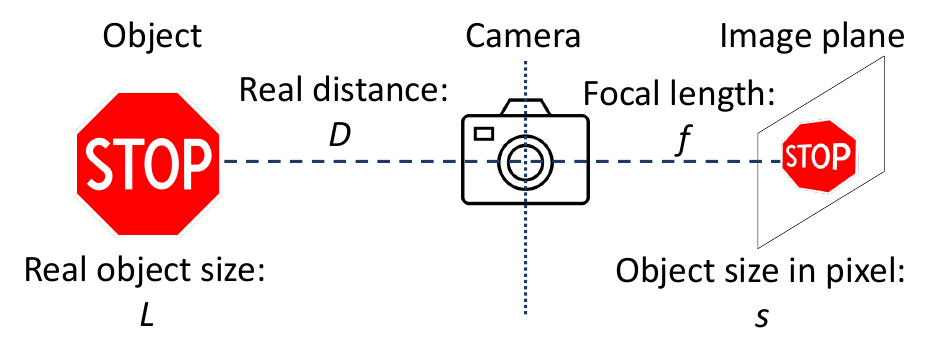}  
          \vspace{-0.7cm}
    \caption{Theoretical analysis of~\S\ref{sec:dlh1}, i.e., the camera pin-hole model.}
        \label{fig:lh1_theory}
\end{minipage}
\vspace{-0.4cm}
\end{figure*}

\textbf{Experimental analysis.} With the same setup in~\S\ref{sec:measurement}, we simulate the real driving scenario in SVL. The STOP sign size in pixels and the distance between the vehicle and the STOP sign can be directly obtained from SVL (\S\ref{sec:measurement-methodology-and-setup}). With that, we can get the frequency distribution histogram over different STOP sign sizes in pixels as shown in the Fig.~\ref{fig:distribution} (b), in which the AD vehicle runs for 30 rounds at speed 25 mph. The distribution shown in Fig.~\ref{fig:distribution} (b) is not uniform, which is wrongly assumed by the prior works~\cite{chen2018shapeshifter, jia2022fooling}. To compare, we sample the STOP sign size from the most recent prior work~\cite{jia2022fooling}, which designs an algorithm to determine the STOP sign size in a \textit{uniform} way. We run that algorithm 30k times and collect the STOP sign size shown in Fig.~\ref{fig:distribution} (a). 
The difference between Fig.~\ref{fig:distribution} (a) and (b) indicates that our observation is held experimentally.

\textbf{Theoretical analysis.} Assuming a uniform motion for AD vehicles, we leverage the camera pin-hole model (Fig.~\ref{fig:lh1_theory}) for the theoretical analysis. From Fig.~\ref{fig:lh1_theory}, we abstract the relationship of real object size ($L$), real distance($D$), focal length($f$), and object size in pixel($s$) with similar triangles: $\frac{L}{D} = \frac{s}{f}$. With the system model shown in Fig.~\ref{fig:system_model}, we assume that the initial vehicle to STOP sign distance is the road length $D_0$ and the current vehicle to STOP sign distance is $D$. Due to uniform motion, the vehicle traveled distance can be formulated as $D_1 = v * t$, where $v$ is the vehicle speed (usually it is the speed limit) and $t$ is the time.
To build the relationship between $s$ and the sampled frequency (i.e., the frame number) $F$, we formulated the time $t$ as $t = \frac{F}{\eta}$, where the $\eta$ is the image capturing frequency from the camera. Due to $D_1 + D = D_0$ and the camera pin-hole model (Fig.~\ref{fig:lh1_theory}), we can obtain the following equation:
\begin{equation}
\label{eq:distance}
\small
D_0 = D + v * \frac{F}{\eta} = \frac{L * f}{s} + v * \frac{F}{\eta} \rightarrow F = (D_0 - \frac{L * f}{s}) * \frac{\eta}{v}
\end{equation}
\meq{eq:distance} is the CDF of $s$, since the $F$ is accumulated frames. 
To obtain PDF, CDF's derivative is calculated:
\begin{equation}
\label{eq:PDF-distance}
\small
F' = \frac{d F}{d s} = \frac{\eta * L * f}{v * s^2}
\end{equation}
From \meq{eq:PDF-distance}, the probability distribution is definitely not uniform.
We also plot \meq{eq:PDF-distance} as shown in Fig.~\ref{fig:distribution} (c) with $\eta = 20$, $L = 1.5$, $v = 25 mph$, and $f = 25 mm$ (commonly used in AD system such as Baidu Apollo). The distribution is similar to the distribution in the experimental analysis shown in Fig.~\ref{fig:distribution} (b), which supports our observation. 

\textbf{Our system-driven solution (S1).} With that, we propose our system-driven solution (S1) to address this inconsistency above. Leveraging the system model in Fig.~\ref{fig:system_model}, we define a novel object size distribution based on \meq{eq:PDF-distance}: $\mathcal{S} = \{s_1, s_2, ..., s_N\}$ as a discrete distribution, where $s_i$ is the object size in pixels. Based on the \meq{eq:PDF-distance}, the probability of $s_i$ can be abstract as $p(s_i) = \frac{1}{s_i^2}/\sum_{k = 1}^{N} \frac{1}{s_k^2}$. Such new object size distribution can be used to address the inconsistency observation and easily integrated into the attack design (\meq{eq:eot-distribution}). However, to get the detailed distribution, we have to know the range of $\mathcal{S}$, which will be addressed in the following system-driven solution (S2) in~\S\ref{sec:dlh2}.

\subsection{Lack of Vehicle Plant Model and AD System Model Consideration}
\label{sec:dlh2}
In the EoT process, uniformly sampling the object size ($\mathcal{S}$ in \meq{eq:eot-distribution}) in a \textit{range} is generally used.
In the prior works, they just treat it as hyper-parameters without any reasons~\cite{chen2018shapeshifter, jia2022fooling}. 
In practice, not every range is equivalently important to achieve system-level effects. Taking STOP sign case as an example, within $d_{min}$ in Fig~\ref{fig:system_model}, despite applying maximum deceleration, AD vehicle still cannot fully stop before the stop line. Thereby, such a range is not important to achieve system-level effects. 
However, none of 
the prior works involve the system-critical range related to \textit{the vehicle plant model} and \textit{AD system model} in their attack designs, which leads to the second observation: \textit{lack of vehicle plant model and AD system model consideration.}

Note that previous studies which indiscriminately utilize a broad range of object sizes for attacks have exhibited reduced effectiveness in comparison to those employing a small size range.
For example, when the object size is small (implying the AD vehicle is far away from the object), attack convergence becomes challenging~\cite{jia2022fooling}, which indicates that it is harder to attack. Therefore, generally, utilizing a more optimally defined range, as opposed to an excessively broad one, enhances the efficacy of the attack.

To elucidate the disparity in attack effectiveness between utilizing a broad versus a narrow object size range, we conducted experiments comparing the attack success rates with small and large range of the STOP sign size. We follow a similar evaluation setup as in~\S\ref{sec:measurement-methodology-and-setup} but use a pure simulation-based setup for RP$_2$ attack.
Specifically, the small range for the STOP sign spans from 30 px to 100 px, whereas the large range extends from 30 px to 416 px, representing the maximum range at which the benign STOP sign is detectable. Results presented in Table~\ref{tab:res_large_range} reveal a superior average attack success rate for the small range over the large range. Although the large range demonstrates promising convergence at close distances, its performance diminishes between 5 to 30 m. This suggests that simply opting for a larger range does not guarantee enhanced performance.

\begin{table}[t]
\tabcolsep 0.015in

\footnotesize
    \caption{Attack success rate of RP$_2$ for Y2 evaluated in simulation with both small and large STOP sign pixel sizes.}
    \vspace{-0.2cm}
    \centering
    \begin{tabular}{cccccccccc}
    \toprule
    & & \multicolumn{7}{c}{Distance (m)} &\\
    \cmidrule(lr){3-9}
      & & 4 - 5 & 5 - 10 & 10 - 15 & 15 - 20 & 20 - 25 & 25 - 30 & 30 - 35 & \multirow{-2}{*}{Ave}\\
     \cmidrule(lr){1-10}
     & Small &  6.7\% & 37.1\% & 68.3\% & 81.1\% & 100\% & 100\% & 100\% & 70.5\% \\
     \multirow{-2}{*}{\shortstack{Comp. \\ ASR}} & Large &  98.6\% & 6.1\% & 0\% & 1.0\% & 58.5\% & 99.1\% & 100\% & 51.9\%\\

        \bottomrule
         
    \end{tabular}

    \label{tab:res_large_range}
    \vspace{-0.5cm}
\end{table}

\textbf{Our system-driven solution (S2).}
We introduce our Solution (S2) to ascertain the system-critical range from the \textit{vehicle plant model} and the \textit{AD system model}. With these, we directly deduce the $d_{min}$ and $d_{max}$ values as shown in Fig.~\ref{fig:system_model}.
Then, we convert these distances to the corresponding object sizes in pixels ($\mathcal{S}$).
From the system model (Fig.\ref{fig:system_model}), it's evident that the minimum braking distance can be used as $d_{min}$. Within this distance, detection results have a negligible impact on system-level effects. As for the $d_{max}$, several tasks in the AD system, such as object detection and tracking, can influence its determination. For object detection, the maximum distance can be the furthest benign distance where an object is detected. For object tracking, we select a conservative tracking~\cite{jia2020fooling} since attackers might not always access the precise tracking parameters of the targeted AD system and a conservative tracking provides a broader system-critical range generally. To achieve system-level effects, the object should not be tracked when the vehicle reaches the $d_{min}$. 
Due to conservative tracking,
such tracking distance (i.e., if within this distance, the object can never be detected, the tracker will be deleted) usually exceeds the distance where the object detector can detect the object. Thus, simplifying this, we select the distance where the benign object can be detected with a small detection rate as $d_{max}$. 
Having deduced the $d_{min}$ and $d_{max}$, the next step involves translating these distances into pixel object sizes ($\mathcal{S}$) and determining the appropriate object location in the background (function $M(.)$ in~\S\ref{sec:attack_design_overview}). We suggest two methodologies to solve address it: 1) camera-based rendering~\cite{xiao2019meshadv, caesar2020nuscenes, sp:2021:ningfei:msf-adv} and 2) manual annotation~\cite{xu2020adversarial}. Employing these methods allows us to acquire precise specifications about position and pixel size range (in~\S\ref{sec:attack-design-eval} and ~\S\ref{sec:general_object_type}).

%% file: src/evaluation.tex
\begin{table*}[t]
\tabcolsep 0.09in
\footnotesize
    \caption{Attack success rates of RP$_2$, FTE-Y3, and FTE-Y5 on STOP sign-evasion attack (\S\ref{sec:hypothesis-val}) and ADV-Tshirt on pedestrian-evasion attack (\S\ref{sec:general_object_type}) for our attack design evaluation with perception results modeling from physical world. + S1: with S1 only; + S2: with S2 only; + S1 + S2: with S1 and S2; + S1 + S2 (TV): + S1 + S2 with TV loss (\S\ref{sec:attack-design-eval}). \textbf{Bolded} numbers indicate the cases where our design outperforms the original baseline attack (``Original'') within the system-critical range.}
     \vspace{-0.2cm}
    \centering
    \begin{tabular}{cclccccccccc}
    \toprule
    & & & \multicolumn{9}{c}{Distance (m): Gray color means the attack success rate within the system-critical range}\\
    \cmidrule(lr){4-12}
     Object detector & \multicolumn{2}{c}{Attack design} & 4 - 5 & 5 - 10 & 10 - 15 & 15 - 20 & 20 - 25 & 25 - 30 & 30 - 35 & 35 - 40 & 40 - 45\\
     \cmidrule(lr){1-12}
     
     & \multicolumn{2}{c}{Original~\cite{eykholt2018physical}} & 41.8\% &  \cellcolor[gray]{0.9} 10.0\% & \cellcolor[gray]{0.9}23.8\% & \cellcolor[gray]{0.9}65.4\% & 99.9\% & 100\% & 100\% & 100\% & 100\% \\
     & &+ S1 & 4.4\% &  \cellcolor[gray]{0.9}\textbf{13.7\%} & \cellcolor[gray]{0.9}\textbf{51.2\% }& \cellcolor[gray]{0.9}\textbf{99.3\%} & 100\% & 100\% & 100\% & 100\% & 100\% \\
     & &+ S2 & 5.6\% & \cellcolor[gray]{0.9}\textbf{44.9\%} & \cellcolor[gray]{0.9}\textbf{57.8\%} & \cellcolor[gray]{0.9}\textbf{98.7\%} & 100\% & 100\% & 100\% & 100\% & 100\% \\
     \multirow{-4}{*}{YOLO v2 (Y2)}& \multirow{-3}{*}{RP$_2$} & + S1 + S2& 36.1\% & \cellcolor[gray]{0.9}\textbf{65.8\%} & \cellcolor[gray]{0.9}\textbf{88.0\%} & \cellcolor[gray]{0.9}\textbf{100\%} & 100\% & 100\% & 100\% & 100\% & 100\%\\
     \midrule
     &\multicolumn{2}{c}{Original~\cite{jia2022fooling}} & 10.1\% & \cellcolor[gray]{0.9}0\% & \cellcolor[gray]{0.9}0\% & \cellcolor[gray]{0.9}12.7\% & \cellcolor[gray]{0.9}57.1\% & \cellcolor[gray]{0.9}99.4\% & 100\% & 100\% & 100\%\\ 
    & &+ S1 & 0\% & \cellcolor[gray]{0.9}0\% & \cellcolor[gray]{0.9}0\% & \cellcolor[gray]{0.9}\textbf{14.0\%} & \cellcolor[gray]{0.9}\textbf{72.2\%} & \cellcolor[gray]{0.9}95.9\% & 100\% & 100\% & 100\%\\ 
      & &+ S2 & 0\% & \cellcolor[gray]{0.9}0\% & \cellcolor[gray]{0.9}0\% &\cellcolor[gray]{0.9}\textbf{13.4\%} & \cellcolor[gray]{0.9}\textbf{81.4\%} & \cellcolor[gray]{0.9}94.4\% & 97.2\% & 100\% & 100\%\\
    \multirow{-4}{*}{YOLO v3 (Y3)} & \multirow{-3}{*}{FTE-Y3} & + S1 + S2 & 5.3\% & \cellcolor[gray]{0.9}0\% & \cellcolor[gray]{0.9}\textbf{34.7\%} & \cellcolor[gray]{0.9}\textbf{94.0\%} & \cellcolor[gray]{0.9}\textbf{99.4\%} & \cellcolor[gray]{0.9}\textbf{100\% }& 100\% & 100\% & 100\% \\
     \midrule
       & \multicolumn{2}{c}{Original~\cite{jia2022fooling}} & 8.8\% & \cellcolor[gray]{0.9}0\% & \cellcolor[gray]{0.9}0\% & \cellcolor[gray]{0.9}0.3\% & \cellcolor[gray]{0.9}11.8\% & \cellcolor[gray]{0.9}51.6\% & \cellcolor[gray]{0.9}96.1\% & 100\% & 100\%\\
      & &+ S1 & 0.3\% & \cellcolor[gray]{0.9}0\% & \cellcolor[gray]{0.9}0\% & \cellcolor[gray]{0.9}\textbf{1.3\%} & \cellcolor[gray]{0.9}\textbf{13.9\%} & \cellcolor[gray]{0.9}\textbf{69.3\%} & \cellcolor[gray]{0.9}94.1\% & 99.0\% & 100\%\\
    & &+ S2 & 1.5\% & \cellcolor[gray]{0.9}0\% & \cellcolor[gray]{0.9}\textbf{0.1\%} & \cellcolor[gray]{0.9}\textbf{1.7\%} & \cellcolor[gray]{0.9}\textbf{32.7\%} & \cellcolor[gray]{0.9}\textbf{81.9\%} & \cellcolor[gray]{0.9}\textbf{99.0\%} & 100\% & 100\%\\
     & &+ S1 + S2 &  16.5\% & \cellcolor[gray]{0.9}0\% & \cellcolor[gray]{0.9}\textbf{4.3\%} & \cellcolor[gray]{0.9}\textbf{47.2\%} & \cellcolor[gray]{0.9}\textbf{93.4\%} & \cellcolor[gray]{0.9}\textbf{99.7\%} & \cellcolor[gray]{0.9}\textbf{100\%} & 100\% & 100\%\\
    \multirow{-5}{*}{YOLO v5 (Y5)}& \multirow{-4}{*}{FTE-Y5} &+ S1 + S2 (TV) &  43.6\% & \cellcolor[gray]{0.9}\textbf{51.7\%} & \cellcolor[gray]{0.9}\textbf{42.1\%} & \cellcolor[gray]{0.9}\textbf{26.3\%} & \cellcolor[gray]{0.9}\textbf{23.8\%} & \cellcolor[gray]{0.9}\textbf{66.1\%} & \cellcolor[gray]{0.9}\textbf{97.7\%} & 99.7\% & 100\%\\
     
    \midrule
    \midrule
     & \multicolumn{2}{c}{Original~\cite{xu2020adversarial}} & 13.5\% & \cellcolor[gray]{0.9}0\% & \cellcolor[gray]{0.9}31.3\% & \cellcolor[gray]{0.9}86.1\% & 96.1\% & 90.7\% & 86.0\% & 100\% & 100\% \\
     \multirow{-2}{*}{YOLO v2 (Y2)}&\multicolumn{2}{c}{ADV-Tshirt + S1 + S2}& 0\% & \cellcolor[gray]{0.9}0\% & \cellcolor[gray]{0.9}\textbf{34.2\%} & \cellcolor[gray]{0.9}\textbf{89.0\%} & 91.7\% & 83.5\% & 78.5\% & 98.7\% & 100\%\\
     \cmidrule(lr){2-12}

     & \multicolumn{2}{c}{Original~\cite{xu2020adversarial}} & 3.8\% & \cellcolor[gray]{0.9}0\% & \cellcolor[gray]{0.9}3.8\% & \cellcolor[gray]{0.9}32.2\% & \cellcolor[gray]{0.9}75.7\% & \cellcolor[gray]{0.9}89.8\% & 90.5\% & 91.5\% & 95.1\% \\
     \multirow{-2}{*}{YOLO v3 (Y3)}&\multicolumn{2}{c}{ADV-Tshirt + S1 + S2}& 0\% & \cellcolor[gray]{0.9}0\% & \cellcolor[gray]{0.9}\textbf{33.6\%} & \cellcolor[gray]{0.9}\textbf{88.2\%} & \cellcolor[gray]{0.9}\textbf{91.3\%} & \cellcolor[gray]{0.9}\textbf{92.4\%} & 89.7\% & 90.7\% & 87.7\%\\
    \cmidrule(lr){2-12}

     & \multicolumn{2}{c}{Original~\cite{xu2020adversarial}} & 35.9\% & \cellcolor[gray]{0.9}6.8\% & \cellcolor[gray]{0.9}17.1\% & \cellcolor[gray]{0.9}36.7\% & \cellcolor[gray]{0.9}37.4\% & \cellcolor[gray]{0.9}72.0\% & \cellcolor[gray]{0.9}88.6\% & 92.3\% & 91.5\% \\
     \multirow{-2}{*}{YOLO v5 (Y5)}& \multicolumn{2}{c}{ADV-Tshirt + S1 + S2}& 2.6\% & \cellcolor[gray]{0.9}1.0\% & \cellcolor[gray]{0.9}\textbf{61.6\%} & \cellcolor[gray]{0.9}\textbf{74.7\%} & \cellcolor[gray]{0.9}\textbf{58.3\%} & \cellcolor[gray]{0.9}\textbf{89.6\%} & \cellcolor[gray]{0.9}\textbf{90.5\%} & 64.1\% & 61.1\%\\

        \bottomrule
         
    \end{tabular}

    \label{tab:benign-modeling}

\end{table*}

\section{Evaluation}
\label{sec:hypothesis-val}
\label{sec:generality_eval}
We adopt the same evaluation methodology and setup as~\S\ref{sec:measurement-methodology-and-setup}.
The printed STOP signs with the newly generated patches
are in Fig.~\ref{fig:stop_sign_attack_visual}.
We evaluate some attacks on one-stage object detectors, i.e. Y2, Y3, and Y5 due to their better real-time performance compared to two-stage ones~\cite{zou2023object}. RP$_2$ and FTE are selected as the evaluated attacks.
Attack generality is evaluated in~\S\ref{sec:general_ad_settings} and~\S\ref{sec:general_object_type}.
The combination for attacks and object detectors are
RP$_2$, FTE-Y3, and FTE-Y5.

\begin{table*}[t]
\tabcolsep 0.02in
\footnotesize
    \caption{System-level violation rate tested in simulation and component-level ASR evaluation including baseline comparison (i.e., Original and ablation studies). Each cell contains 10 runs with different initial positions of the AD vehicle. S1: with S1 only; S2: with S2 only; S1 + S2: with S1 and S2; SCR: System-critical range (\S\ref{sec:dlh2}). $^*$ with special improvements (\S\ref{sec:attack-design-eval}).}
    \vspace{-0.2cm}
    \centering
    \begin{tabular}{cccccccccccccc}

    \toprule
  &  & \multicolumn{4}{c}{RP$_2$} & \multicolumn{4}{c}{FTE-Y3} & \multicolumn{4}{c}{FTE-Y5}\\
    
    \cmidrule(lr){3-6}
    \cmidrule(lr){7-10}
    \cmidrule(lr){11-14}
    
  \multirow{-2}{*}[3pt]{\shortstack{Evaluation level}} & \multirow{-2}{*}[3pt]{\shortstack{Speed (mph)}}& Original~\cite{eykholt2018physical} & S1 & S2 & S1 + S2 & Original~\cite{jia2022fooling} & S1 & S2 & S1 + S2 &Original~\cite{jia2022fooling} & S1 & S2 & S1 + S2 \\
    
    \midrule

  &  25 & 0\% &  90\% &   100\% & 100\% & 0\% & 0\% & 0\% & 40\%& 0\% &0\% & 0\%& 10\%$^*$ \\
  &  30 & - &  - &   - & - &0\% & 0\% &30\%  & 100\%& 0\% & 0\% & 0\% &  80\%\\
    \multirow{-3}{*}{\shortstack{System (violation rate)}} &35 & - & - & - & - & - & - & - & - & 0\%& 30\%  &  40\% & 100\%\\
    
    \midrule
    \multicolumn{2}{c}{$p$-value} & - & 0.00 & 0.00 & 0.00 & -& - & 0.08 & 0.00 & -& 0.08 & 0.04 & 0.00\\
    \midrule
    & Overall & 71.2\% & 74.2\% & 78.6\%  & 87.8\% & 53.3\%& 53.6\%& 54.0\% & 70.4\% & 41.0\% &42.0\% &46.3\% &62.3\%\\
\multirow{-2}{*}{\shortstack{Component (ASR)}}& \shortstack{SCR} & 33.1\% & 54.7\%  & 67.1\% & 84.6\% & 33.8\%& 36.4\%& 37.8\% &65.6\% &26.6\% &29.8\% &35.9\% &57.4\%\\

        \bottomrule

    \end{tabular}
\vspace{-0.3cm}
    \label{tab:violation-our-work}

\end{table*}

\subsection{System-level Attack Effectiveness Evaluation}
\label{sec:attack-design-eval}
{\bf Attack generation.}
We adopt the attack methodology in~\S\ref{sec:design-lh-improve-proposal}.
We employ a camera-based rendering method and utilize the nuScenes dataset~\cite{caesar2020nuscenes} to translate the system-critical range from the physical world to the pixel range in images. Notably, nuScenes offers APIs that facilitate rendering objects within images.
Specially, we render the four corners of the STOP sign and obtain its size in pixels by measuring the distance between these four corner points in the image. With S1 and S2, we can embed the system-model property into the attack generation process to improve system-level effects.
To further validate the effects of S1 and S2, we perform ablation studies by generating the attack with S1 only and S2 only, and comparing them to the attack generated with/without both S1 and S2. Details of attacks without S1 
and S2 (i.e., original attacks) are in~\S\ref{sec:measurement}.

\begin{figure}[t]
    \footnotesize
      \centering
          \includegraphics[width=\linewidth]{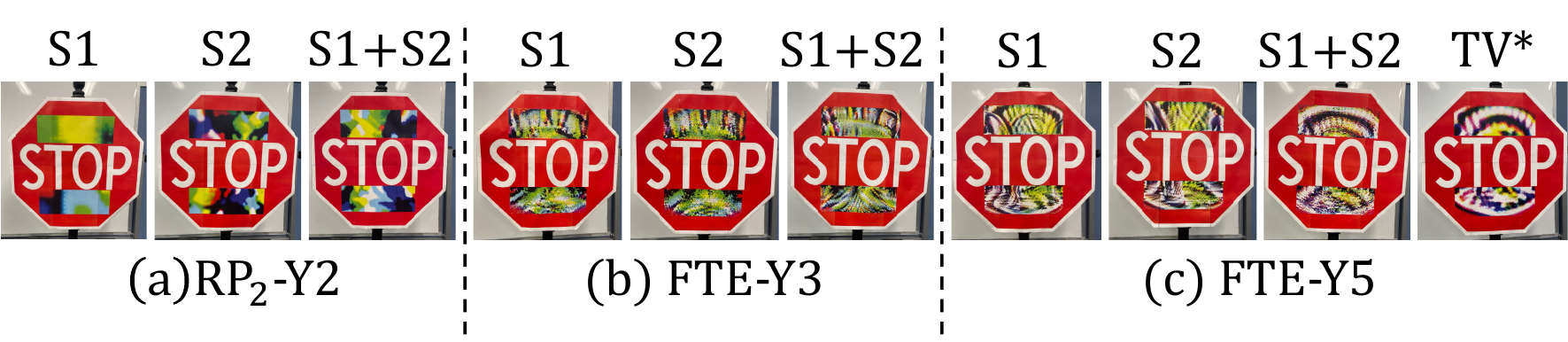}  
         \vspace{-0.6cm}
    \caption{Visualization of STOP sign attacks with system-driven design. S1: with S1 only; S2: with S2 only; S1 + S2: with both S1 and S2; TV$^*$: S1 + S2 with TV loss (\S\ref{sec:attack-design-eval}). }
    \vspace{-0.4cm}
        \label{fig:stop_sign_attack_visual}
\end{figure}

{\bf Results.} The STOP sign attack images are in Fig.~\ref{fig:stop_sign_attack_visual}, which are printed in physical world and the perception modeling results from the physical world are in Table~\ref{tab:benign-modeling}. From the results in Table~\ref{tab:benign-modeling}, almost all the results (bolded in the table) with our system-driven attack improvement can outperform the original attack. As shown in Table~\ref{tab:violation-our-work}, with our system-driven attack designs, the system-level violation rate can increase by around 70\% on average, where we only include the results where the benign cases have a 0\% system-level violation rate. The $p$-value (Table~\ref{tab:violation-our-work}) is generally at the statistically significant level (e.g., generally $<$ 0.05 or at a similar magnitude, especially for S1+S2). With S1 + S2, the overall component attack success rate can increase by around 33\% on average. Especially, in the system critical range, the attack success rate can increase by 122\%, which can significantly improve the system-level effects.
Taking FTE-Y5 at 35 mph as an example, the brake distance of 35 mph is around 20 m and the attack success rate from 20 - 35 m shown in Table~\ref{tab:benign-modeling} is around 98\%, which shows a high chance to make the STOP sign not tracked before the brake distance, which leads to the 100\% violation rate (Table~\ref{tab:violation-our-work}). 

For FTE-Y5 at 25 mph, due to the low effectiveness (i.e., around 4\%) from 10 m to 15 m, the tracker cannot be deleted, which leads to 0\% system violation. Thus, we provide a \textit{special improvement} by applying the total variation (TV) loss as prior works~\cite{eykholt2018physical, xiao2019meshadv} which benefits the attack effectiveness. The perception modeling results from the physical world are in Table~\ref{tab:benign-modeling} and the attack visualization is shown in Fig.~\ref{fig:stop_sign_attack_visual}. The system violation rate increases to 10\% after improvement as shown in Table~\ref{tab:violation-our-work} with $^*$. Based on results in Table~\ref{tab:benign-modeling}, the attack success rate in a near distance is generally lower, which aligns well with the results of prior work~\cite{zhao2019seeing}. This leaves space for future works: improving component attack success rate in the near distance.

The results of the ablation study are also summarized in Table~\ref{tab:violation-our-work}. Although in the majority of cases, S1 cannot significantly improve the system-level effects (20\% on average), the component attack success rate in the system-critical range is improved. Compared to S1, S2 has better results (around 28\% on average).
Only combining S1 and S2 can further benefit the system-level effects 
(around 70\% on average), which shows the necessity of both S1 and S2.

%% file: src/discussion.tex
\begin{table}[t]
\tabcolsep 0.04in
\footnotesize
    \caption{System-level violation rate tested in simulation on different AD parameter settings which are highly critical to the system-level effects. The perception modeling results from physical world are in Table~\ref{tab:measurement-results} and Table~\ref{tab:violation-our-work}. }
    \vspace{-0.2cm}
    
    \centering
    \begin{tabular}{ccccccc}

    \toprule
    Tracking param ($H$, $R$) & \multicolumn{2}{c}{(4, 6)~\cite{Apollo2022baidu}} & \multicolumn{2}{c}{ (3, 5)~\cite{kato2018autoware}} & \multicolumn{2}{c}{ (4, 40)~\cite{zhu2018online}} \\
    \cmidrule(lr){2-3}
    \cmidrule(lr){4-5}
    \cmidrule(lr){6-7}
    Brake ($m/s^2$) & -3.4 & -6.0 & -3.4 & -6.0 & -3.4 & -6.0 \\
    \midrule
    Original~\cite{jia2022fooling} & 20\% & 0\% & 50\% & 0\% & 40\% & 0\%\\
    \textbf{Ours} & \textbf{100\%} &\textbf{100\%} & \textbf{100\%} & \textbf{90\%}& \textbf{100\%} & \textbf{100\%}\\
    \bottomrule
    
    \end{tabular}

    \label{tab:general-ad-settings}
\vspace{-0.3cm}
\end{table}

\subsection{Generality on Different AD System Parameters} 
\label{sec:general_ad_settings}

{\bf Methodology and setup.} We select the most safety-critical parameters on system-level effects in AD systems for this evaluation including the tracking parameters ($H$, $R$), where the tracking creates a tracker for an object only when it is continuously detected for $H$ frames, and deletes its tracker only when the object continuously disappears for $R$ frames~\cite{zhu2018distractor, Apollo2022baidu, kato2018autoware, jia2020fooling}, and the brake deceleration where we use the safe
vehicle deceleration and max vehicle deceleration~\cite{brake-accele}. We select the tracking parameters from 
Baidu Apollo~\cite{Apollo2022baidu} and Autoware.AI~\cite{kato2018autoware}, and the representative research paper~\cite{zhu2018online}. All the parameter details are in Table~\ref{tab:general-ad-settings} and for others, we follow the same setup in~\S\ref{sec:attack-design-eval}. We select the FTE-Y5 since it is the most representative attack so far and 35 mph as the target speed due to its high safety impact.

{\bf Results.} The system-level attack effect results (violation rate) are summarized in Table~\ref{tab:general-ad-settings}, where we compared our attack with the original naive attacks (\S\ref{sec:measurement}). The results show that our attack can outperform the original attack in all the different AD parameter settings on the system-level effect. On average, we have around 98\% system violation rate (5 times larger than the original one) while the original naive attack only has 18\%. The results further point out that our attack is general to different critical AD system parameters.

\begin{table}[t]
\tabcolsep 0.03in
\footnotesize
    \caption{Pedestrian collision rate tested in simulation with ADV-Tshirt attack on different object detectors. 10 runs for each cell with different initial AD vehicle position.}
    \centering
    \vspace{-0.2cm}
    \begin{tabular}{ccccccc}

    \toprule
    & \multicolumn{2}{c}{YOLO v2 (Y2)} & \multicolumn{2}{c}{YOLO v3 (Y3)} & \multicolumn{2}{c}{YOLO v5 (Y5)} \\
     \cmidrule(lr){2-3}
     \cmidrule(lr){4-5}
     \cmidrule(lr){6-7}
    \multirow{-2}{*}[0.1cm]{\shortstack{Speed \\(mph)}} & Original~\cite{xu2020adversarial} & \textbf{Ours} & Original~\cite{xu2020adversarial} & \textbf{Ours} & Original~\cite{xu2020adversarial} & \textbf{Ours}\\
    \midrule
    25 &  20\% & \textbf{50\%} & 0\% & \textbf{50\%} & 0\% & \textbf{70\%}\\
    30 &  100\% & \textbf{100\%}& 50\% & \textbf{100\%} & 10\% & \textbf{80\%} \\
    35 & 100\%  &\textbf{100\%} & 80\% & \textbf{100\%} & 60\% & \textbf{90\%} \\

        \bottomrule
         
    \end{tabular}

    \label{tab:general-advtshirt}
\vspace{-0.3cm}
\end{table}
\subsection{Generality on a Different Object Types}
\label{sec:general_object_type}

\begin{figure}[t]
    \footnotesize
      \centering
 \includegraphics[width=\linewidth]{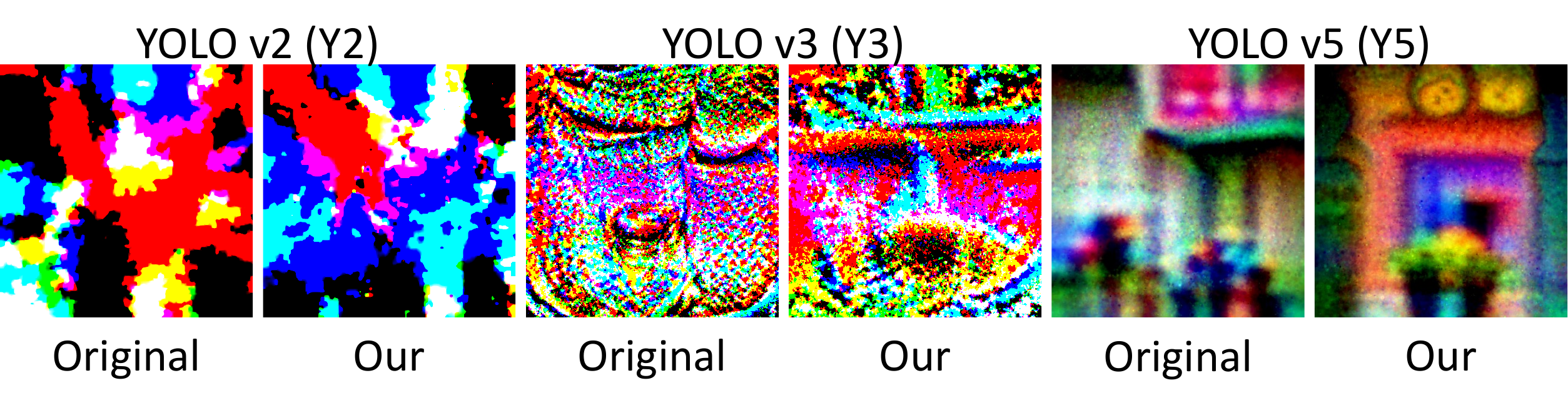}  \vspace{-0.7cm}
    \caption{Visualization of ADV-Tshirt attack with and without system-driven design.}
        \label{fig:adv_patch}
        \vspace{-0.3cm}
\end{figure}

{\bf Methodology and setup.} We select the ``pedestrian'' as our target object type since making the pedestrian vanish will cause a significant impact on AD. We select the most representative patch attack -- adversarial T-shirt~\cite{xu2020adversarial}, which is called ADV-Tshirt in our paper. For object detectors, we select Y2, Y3, and Y5 and follow the same setup in ADV-Tshirt paper~\cite{xu2020adversarial}, and we collect the videos from the real world (similar methodology in~\S\ref{sec:measurement}) for attack generation and manually annotate the four corner points for placing the patch (obtaining the size and position~\S\ref{sec:dlh2}). Each video segment has around 200 frames. 
We perform digital perception result modeling with real-world data we collected.

{\bf Results.} The perception results modeling results for ADV-Tshirt are shown in Table~\ref{tab:benign-modeling} and the generated patches are visualized in Fig.~\ref{fig:adv_patch}. We define the system-level effect metric as pedestrian collision rate: $\dfrac{K_\textrm{collision}}{K_\textrm{total}}$, in which $K_\textrm{collision}$ means the number of runs where the AD vehicle crash into the pedestrian, and $K_\textrm{total}$ is the number of total runs.
The system-level evaluation with the comparison with the original attack~\cite{xu2020adversarial} is shown in Table~\ref{tab:general-advtshirt}. In average, our attack designs can achieve around 82\% pedestrian collision rate while the original attack can only achieve around 47\% pedestrian collision rate. Especially, for the most advanced object detector such as Y3 and Y5, our pedestrian collision rate has significant improvement compared to the original attack.
Y2 is more fragile than others which makes the original attack have very high attack effectiveness in the component level and leads to pedestrian collision rates at the similar level as ours. The results show the generality of our attack designs to different object types which further shows the generality to different system models (\S\ref{sec:background-system-model}).

%% file: src/limitation.tex
\section{Discussion}
\label{sec:discussion}
{\bf Potential mitigation.} 
The ongoing tug-of-war between adversarial attacks and their defenses has yielded a range of mitigation strategies, such as adversarial training~\cite{madry2017towards}. 
Since several object-evasion attacks in AD context have been identified~\cite{zhao2019seeing, jia2022fooling}, there is an immediate need for defense exploration. Before pursuing novel mitigation strategies, it is imperative to first measure how existing defenses affects 
system-level attack effectiveness in AD, especially the ones with theoretical guarantees~\cite{xiang2023objectseeker, xiang2021detectorguard}, which should be a future work. Another promising direction involves cross-checking with alternate perception sources. For example, AD systems might verify camera-based pedestrian detection with LiDAR perception. Despite not offering a fundamental defense strategy~\cite{sp:2021:ningfei:msf-adv}, they may make system-level attack effects more difficult to achieve.
Thus, we leave a systematic exploration of these defenses to future work.

{\bf Limitation and future work.} First, although we leverage the perception results that modeling from the physical world and demonstrate the system-level effects in AD system with LGSVL, the feasibility of the attack effects on real AD systems in physical world remains unclear. Thus, the exploration of the attack practicality is a valuable future work. Second, our attack is within white-box threat model, which is less practical compared to black-box one. Thereby, the development of a novel attack with a practical threat model is a potential future work. Third, although we explore the generality on different AD system parameters in \S\ref{sec:general_ad_settings}, our evaluation results and findings are limited by the current AD system setups introduced in \S\ref{sec:measurement-methodology-and-setup}. Therefore, the system-level effect measurement on commercial AD systems such as Tesla is an important future direction.

%% file: src/conclusion.tex
\section{Conclusion}
In this paper, we ask whether previous works can achieve system-level effects
(e.g., vehicle collisions, traffic rule violations) 
under real AD settings. Then, we perform the first measurement study to answer this research question. Our evaluation results show that all representative prior works cannot achieve any system-level effects in a closed-loop AD setup due to the lack of the system model. With our newly proposed system-driven designs, i.e., \system, the system-level effects can be significantly improved. We hope that the concept of the system model could guide future security analysis/testing for real/practical AD systems.

%% file: src/ack.tex
\section{Acknowledgments}
We would like to thank Ziwen Wan, Junjie Shen, Tong Wu, Junze Liu, Fayzah Alshammari, Trishna Chakraborty, and the anonymous reviewers for their valuable and insightful feedback. 
This research was supported by the NSF under grants CNS-1932464, CNS-1929771, and CNS-2145493.